\documentclass[12pt]{article}
\usepackage{cite,epsfig,amssymb,amsmath,graphicx,color}
\newcommand{\be}{\begin{equation}}
\newcommand{\ee}{\end{equation}}
\newcommand{\bear}{\begin{eqnarray}}
\newcommand{\eear}{\end{eqnarray}}
\newcommand{\ba}{\begin{array}}
\newcommand{\ea}{\end{array}}
\newcommand{\nn}{\nonumber}

\begin{document}

\begin{center}
{{{\Large \bf  Perturbations in Symmetric Lee-Wick Bouncing Universe}
}\\[17mm]
Inyong Cho$^{1}$ and O-Kab Kwon$^{2}$  \\[3mm]
{\it $^{1}$Institute of Convergence Fundamental Studies \& School of Liberal Arts, \\
Seoul National University of Science and Technology,
Seoul 139-743, Korea}\\[2mm]
{\it $^{2}$Department of Physics,~BK21 Physics Research Division,
~Institute of Basic Science,\\
Sungkyunkwan University, Suwon 440-746, Korea},\\[2mm]}
{\tt iycho@seoultech.ac.kr,~okab@skku.edu}
\end{center}

\vspace{10mm}

\begin{abstract}
We investigate the tensor and the scalar perturbations
in the symmetric bouncing universe
driven by one ordinary field
and its Lee-Wick partner field which is a ghost.
We obtain the even- and the odd-mode functions of the tensor perturbation
in the matter-dominated regime.
The tensor perturbation grows in time during the contracting phase of the Universe,
and decays during the expanding phase.
The power spectrum for the tensor perturbation is evaluated
and the spectral index is given by $n_{\rm T} =6$.
We add the analysis on the scalar perturbation
by inspecting the even- and the odd-mode functions in the matter-dominated regime,
which was studied numerically in our previous work.
We conclude that the comoving curvature by the scalar perturbation is constant
in the super-horizon scale and starts to decay in the far sub-horizon scale
while the Universe expands.
\end{abstract}
\newpage

\tableofcontents


\section{Introduction}

Inspired by the original work~\cite{Lee:1969fy} of Lee and Wick
on the quantum electrodynamics with higher-derivative propagators,
Grinstein {\it et al} proposed the Lee-Wick standard model~\cite{Grinstein:2007mp}.
In this model, the minimal set of the higher-derivative field was introduced,
and the corresponding $N=2$ Lee-Wick (LW) formalism was obtained.
The $N=2$ LW theory was extended to the $N=3$ one in Refs.~\cite{Carone:2008iw}.
For the higher-derivative model consisted of one scalar field,
the general LW formalism was constructed in Ref.~\cite{Cho:2010hj}.
(See also Refs.~\cite{Grinstein:2007iz,Fornal:2009xc,Accioly:2011zz}
for recent works for the LW field theory.)

The $N=2$  scalar-field LW  model consists of one ordinary scalar field
and its LW partner field which is a ghost.
When this model is applied to cosmology,
the background universe experiences bouncing
due to the energy-condition violating ghost field.
For the bouncing-universe models, for example the ekpyrotic model~\cite{Khoury:2001wf},
the perturbation becomes singular in general about the bouncing point.
However, very recently the $N=2$ scalar-field LW model
was investigated in Ref.~\cite{Cai:2008qw},
and the authors found that the density perturbation at the bouncing point is nonsingular.
(See also Refs.~\cite{Cai:2009fn}.)
Although the stability of the bouncing universe about the anisotropic perturbation
is still questioned, at the linear level it is believed that the LW bouncing universe
is stable about the perturbation.
(See also Refs.~\cite{Zhang:2010bb}
for other types of nonsingular bouncing universe models.)

In Ref.~\cite{Cho:2011re}, we investigated the scalar perturbation
of the {\it symmetric} Lee-Wick bouncing universe,
in which a new type of initial vacuum solution was discovered.
We discussed the growth of the initial perturbation and the resulting late-time power spectrum
by introducing the even- and odd-mode analysis.
In this paper, we investigate the tensor perturbation
applying the even/odd-mode analysis,
and complete the discussion on the scalar perturbation
which was studied in our previous work.

The paper is organized as following.
In Sec.~2, we review the set-up of the $N=2$ symmetric Lee-Wick model
and the evolution of the background universe.
In Sec.~3, we solve the tensor perturbation, and obtain the power spectrum and the spectral index.
In Sec.~4, we extend the mode analysis to the scalar perturbation,
and comment on the shortness of the analysis.
In Sec.~5, we conclude.

\section{Symmetric Bouncing Universe}\label{symbac}
The $N=2$ Lee-Wick model consists of one ordinary scalar field $\varphi_1$
and one ghost field $\varphi_2$.
The Einstein-Hilbert action with these Lee-Wick matter fields is given by
\begin{align}\label{LWmatter}
S_{{\rm LW}} = \int d^4x \sqrt{-g}\left[ \frac{R}{16\pi G} + \sum_{n=1}^{N=2} (-1)^n
\left(\frac12 \partial_\mu\varphi_n\partial^\mu\varphi_n + \frac12
m_n^2\varphi_n^2\right)\right],
\end{align}
where we assumed no interacting potential between the matter fields.
For $m_2^2 > m_1^2$, the ghost field plays an important role about the bouncing point,
and becomes subdominant for the rest of time.
The metric ansatz for the isotropic flat Friedmann-Robertson-Walker universe is given by
\begin{align}\label{FRW}
ds^2 = -dt^2 + a^2(t) dx^idx^i.
\end{align}
The Friedmann equations and the scalar-field equations are then given by
\begin{align}
&H^2 = \frac{8\pi G}{3}\sum_{n=1}^{2}
(-1)^{n+1}\Big(\frac{1}{2}\dot\varphi_n^2 + \frac12
m_n^2\varphi_n^2\Big),
\label{bgHsq} \\
&\dot H = -4\pi G\sum_{n=1}^{2} (-1)^{n+1} \dot\varphi_n^2,
\label{bgHdot} \\
&\ddot\varphi_n + 3H\dot\varphi_n + m_n^2\varphi_n=0,\qquad
(n=1,2).
\label{bgmeq}
\end{align}
The ghost field $\varphi_2$ violates the null energy condition,
and thus the background evolution can have a bouncing from contracting to expanding.

When the Universe undergoes bouncing (let us set the bouncing moment at $t=0$),
the velocity of the scale factor vanishes, $\dot a(0) =0$,
i.e., the Hubble parameter becomes $H(0) =0$.
From the Friedmann equation \eqref{bgHsq}, we have
\begin{align}
\dot a(0)=0 \quad\Leftrightarrow\quad
\left[ \dot\varphi_1^2 - \dot\varphi_2^2 + m_1^2\varphi_1^2  - m_2^2\varphi_2^2 \right]_{t=0}=0.
\end{align}
For the symmetric bouncing, the scalar field $\varphi_n$ need be
symmetric (even or odd) about $t=0$ as well as the scale factor.
In order for $\varphi_n$ to be symmetric, the necessary condition is
\begin{align}
\varphi_n(0) =0,
\quad\mbox{ or }\quad
\dot\varphi_n(0)=0.
\label{symphi}
\end{align}
There are four cases satisfying the above symmetric condition for $\varphi_n$
as it was studied in Ref.~\cite{Cho:2011re}.
Among them, we selected most conventional conditions for the bouncing universe,
\begin{align}
\dot\varphi_n(0) =0
\quad\mbox{ and }\quad
m_1^2\varphi_1^2(0)  = m_2^2\varphi_2^2(0) \neq 0,
\end{align}
which indicates that $\varphi_n$'s are even functions.
This conditions are depicted in Fig.~1.
At $t=0$ the ordinary field rolls down the potential from rest,
while the ghost field climbs up the potential.
Afterwards, the two fields oscillate about $\varphi_n =0$ behaving as dusts.
Then the background universe undergoes the matter-dominated expansion.
Due to the background expansion, the matter-field oscillations are damped.
Setting the ghost field more massive, $m_1^2<m_2^2$, the late-time evolution
is dominated by the light ordinary field.
The background fields at late times approximate as
\begin{align}\label{BGfields}
a(t) \approx \alpha t^{\frac{2}{3}}, \quad
H(t) \approx \frac{2}{3t}, \quad
\varphi_1(t) \approx \frac{\cos (m_1 t + \phi_1)}{\sqrt{3\pi G}\, m_1t},
\end{align}
where $\alpha$ and $\phi_1$ are a numerical constant.
Since the background fields $a(t)$ and $\varphi_n(t)$ are symmetric about $t=0$,
starting from some initial moment $t=t_i<0$, the scalar fields oscillate about $\varphi_n=0$,
and the scale factor decreases, i.e., the Universe contracts.
Meanwhile, the amplitude of $\varphi_n$ increases and reaches the maximum at $t=0$,
and then $\varphi_n$ evolves as stated above for $t>0$.
(The numerical plot of the background fields are plotted in Fig.~1 in Ref.~\cite{Cho:2011re}.)

We consider linear perturbations in this background.
The initial scalar and tensor perturbations are produced
during the contracting phase at $t = t_i \ll 0$.
The initial perturbation for a given wave number crosses the horizon four times in total.
The horizon scale $|H^{-1}|$ is plotted in Fig.~2.
The initial perturbations are produced deep inside the horizon at $t=t_i$,
and then cross out the horizon.
When the perturbations finally cross in the horizon during the expanding phase at $t>0$,
the power spectrum is evaluated.
Afterwards, they enter the nonlinear regime,
which is beyond our concern in this work.

For the scalar perturbation in the symmetric background universe,
rather than solving the perturbation from $t=t_i$ numerically,
one is allowed to solve from $t=0$ to either direction
owing to the symmetry.
In Ref.~\cite{Cho:2011re}, we analyzed the perturbation about $t=0$,
and found that the perturbation is composed of two linearly independent modes (even and odd modes).
Then we solved numerically the even and the odd mode from $t=0$ in the $t>0$ region,
which is numerically stable.
Due to symmetry, we need not solve the perturbation for $t<0$.
This is the advantage of solving the even and the odd mode
rather than other modes in numerical calculations.

\section{Tensor Perturbation}
In this section, we investigate the tensor perturbation in our symmetric bouncing background.
The tensor perturbation is decoupled from the scalar and the vector ones in the linear perturbation theory.
In order to describe the tensor perturbation, let us consider the perturbed
metric in the conformal time,
\begin{align}\label{tenmet}
ds^2 = a^2(\eta) \left[ -d\eta^2 +(\delta_{ij} + \bar{h}_{ij}) dx^idx^j \right],
\end{align}
where the tensor perturbation $\bar h_{ij}$ satisfies the gauge conditions,
\begin{align}\label{tencons}
\partial_i\bar h_{ij}=0,\quad \bar h_{ii}=0.
\end{align}
With these conditions, the tensor mode $\bar h_{ij}$
possesses two degrees of freedom corresponding to
two polarizations of gravitational waves,
and can be expanded as
\begin{align}\label{hexp}
\bar{h}_{ij}(\eta,\vec{x}) =\sqrt{32\pi G}\sum_{\lambda = +,-}
 \frac{\tilde\mu_{\lambda} (\eta,\vec{x})}{a}
\; \epsilon^{\lambda}_{ij},
\end{align}
where $\epsilon^{\lambda}_{ij}$ represents the polarization tensor.
Inserting \eqref{tenmet} and \eqref{hexp} into the Einstein-Hilbert action,
we obtain the action for the tensor perturbation,
\begin{align}\label{muact}
S_{\rm T} = \frac12\sum_{\lambda = +,-} \int d\eta d^3 x \Big[ \big(\partial_\eta\tilde\mu_\lambda\big)^2 - \big(\partial_i\tilde\mu_\lambda\big)^2
+ \frac{a''}{a}\tilde\mu_\lambda^2\Big],
\end{align}
where $'\equiv d/d\eta$ and we used the relation $\epsilon_{ij}^\lambda\epsilon^{ij}_{\lambda'} = \delta^\lambda_{\lambda'}$.
The general solution to the equation of motion for $\tilde\mu_\lambda$ is given by
\begin{align}
\tilde\mu_\lambda(\eta,\vec{x})= \int \frac{d^3k}{(2\pi)^{3/2}} \;\left[
\mu_\lambda(\eta;k)a_{\vec k} + \mu_\lambda^*(\eta;k)a_{-\vec k}^\dagger\right] e^{i\vec k\cdot\vec x},
\end{align}
where $\mu_\lambda(\eta;k)$ satisfies
\begin{align}\label{tenMS}
\mu_{\lambda}'' + \left(k^2 -\frac{a''}{a}\right)\mu_\lambda =0
\end{align}
with the normalization for a given polarization mode $\lambda $,
\begin{align}\label{ten_norm}
\mu_\lambda\mu_\lambda^{*'} -\mu_\lambda^*\mu_\lambda' = i.
\end{align}

\subsection{Initial and Super-Horizon Scale Perturbations}
Let us investigate the mode of the tensor perturbation,
which is produced in the matter-dominated regime during the contracting phase.
The vacuum states of small comoving-wave numbers (long-wave lengths)
exit the Hubble horizon while the Universe contracts.
After they exit the Hubble horizon, they experience the bouncing phase.
At the bouncing point, the Hubble radius becomes infinity as shown in Fig.~2.
Therefore, all the Fourier modes in the super-horizon scale
re-enter into the Hubble horizon near the bouncing point,
and then re-exit the horizon shortly after bouncing during the expanding phase.

When the background universe is in the matter-dominated regime,
the scale factor becomes $a \approx \alpha t^{2/3}$.
One can transform the cosmological time to the conformal time by $dt = ad\eta$,
then obtain $t=\alpha^3\eta^2/27$ and $a\approx \alpha^3\eta^2 /9$.
Then one has $a''/a \approx 2/\eta^2$,
and the equation \eqref{tenMS} becomes
\begin{align}\label{ten_matter}
\mu_{\lambda}'' + \left(k^2 -\frac{2}{\eta^2}\right)\mu_\lambda \approx 0.
\end{align}
This equation looks the same with that of the single-field inflationary models,
but the perturbation described by the tensor mode $\bar h_{ij}$ is different
because the scale fact $a$ is different.

The general solution to Eq.~\eqref{ten_matter} is given by
\begin{align}
\mu_{\lambda}(\eta,k) &= \sqrt{\frac{-\pi k\eta}{2}}
\left[ B_\lambda (k) H^{(1)}_{\frac32}(-k\eta)
+ A_\lambda(k) H^{(2)}_{\frac32} (-k\eta)\right],
\\
&= B_\lambda (k) e^{-ik\eta}\left( -1 + \frac{i}{k\eta}\right)
+ A_\lambda(k) e^{ik\eta} \left(-1 -\frac{i}{ik\eta}\right) \label{negmu}
\\
&\equiv \mu_\lambda^{\rm Re}(\eta) + i \mu_\lambda^{\rm Im}(\eta),
\end{align}
where $H_n^{(1,2)}$ represents the Hankel functions,
and $|B_\lambda|^2 - |A_\lambda|^2 = 1/2k$ from the normalization \eqref{ten_norm}.
If we introduce the even and the odd mode which are linearly independent {\it real} functions,
\begin{align}\label{muEO}
\mu_\lambda^{{\rm E}} \equiv  \cos(k\eta) - \frac{\sin(k\eta)}{k\eta} ,
\qquad
\mu_\lambda^{{\rm O}}\equiv  \sin(k\eta) + \frac{\cos(k\eta)}{k\eta} ,
\end{align}
$\mu_\lambda^{\rm Re}$ and $\mu_\lambda^{\rm Im}$ are expressed by a linear combination of
these even and odd modes.
By setting $A_\lambda = \alpha^\lambda_1 +i\alpha^\lambda_2$ and $B_\lambda = \beta^\lambda_1 +i\beta^\lambda_2$,
one obtains the following relation,
\begin{align}
\left(
\begin{array}{c}
  \mu^{\rm Re}_\lambda \\
 \mu^{\rm Im}_\lambda
\end{array}
\right)
=
\left(
\begin{array}{cc}
  -\alpha^\lambda_1-\beta^\lambda_1 & \alpha^\lambda_2 - \beta^\lambda_2 \\
  -\alpha^\lambda_2 -\beta^\lambda_2 & -\alpha^\lambda_1+\beta^\lambda_1
\end{array}
\right)\left(
\begin{array}{c}
  \mu^{\rm E}_\lambda \\
 \mu^{\rm O}_\lambda
\end{array}
\right).
\label{muTR}
\end{align}

\vspace{12pt}
\noindent \textbf{(i) Initial Perturbation: }
The initial perturbation is produced deep inside the horizon
when the Universe contracts.
Then the perturbation crosses out the horizon
when its physical wave-length scale is comparable with the horizon scale,
\begin{align}
\lambda_{ph} = \frac{a}{k} \sim  \left| H^{-1} \right|
\quad\Longrightarrow\quad
|k\eta| \sim 2.
\end{align}
In the far sub-horizon limit, $|k\eta| \gg 2$,
the even and the odd modes are approximated by
\begin{align}\label{muapprox}
\mu_\lambda^{{\rm E}} \approx  \cos(k\eta),
\qquad
\mu_\lambda^{{\rm O}} \approx  \sin(k\eta).
\end{align}
We adopt the Bunch-Davies vacuum for the production of the
initial perturbation
by taking only the positive-energy mode, $A_\lambda=0$  ($\alpha^\lambda_1=\alpha^\lambda_2=0$).
Then using Eq.~\eqref{muapprox}, or directly from Eq.~\eqref{negmu},
we obtain the initial perturbation,
\begin{align}\label{inivac}
\mu_\lambda  \approx -B_\lambda e^{-ik\eta}
= |B_\lambda|e^{i\theta}e^{-ik\eta} = \frac{1}{\sqrt{2k}}e^{i\theta}e^{-ik\eta},
\end{align}
where $\theta=\tan^{-1}(\beta^\lambda_2/\beta^\lambda_1)$ is a phase factor from the complexity of $B_\lambda$.
The amplitude of the tensor mode becomes
\begin{align}
| \bar h_{ij} | \propto \frac{|\mu_\lambda|}{a} \propto \frac{1}{\eta^2},
\label{hamp}
\end{align}
so it grows initially in the contracting universe.

\vspace{12pt}
\noindent \textbf{(ii) Super-Horizon Scale Perturbation and Power Spectrum: }
Let us discuss the perturbation in the super-horizon scale during the expanding phase,
and its horizon crossing.
We have $|k\eta|\ll 2$ in the super-horizon scale, and
$|k\eta|\sim 2$ at the horizon crossing.
Therefore, the second term in $\mu_\lambda^{{\rm E,O}}$ in Eq.~\eqref{muEO}
is dominant, or comparable to the first term.
The solutions  $\mu_\lambda^{{\rm E}}$ and $\mu_\lambda^{{\rm O}}$ in Eq.~\eqref{muEO}
are valid once the background is matter-dominated.
Therefore, the conditions imposed for the initial perturbation
such as the Bunch-Davies vacuum and normalization,
are still valid in the super-horizon scales,
\begin{align}\label{coefAB2}
A_\lambda =0\,\Leftrightarrow\, \alpha^\lambda_1 = \alpha^\lambda_2 =0,
\qquad\mbox{ and }\qquad
|B_\lambda|^2 = \left(\beta^\lambda_1\right)^2 + \left(\beta^\lambda_2\right)^2 = \frac1{2k}.
\end{align}
Then from the solutions $\mu_\lambda^{{\rm E}}$ and $\mu_\lambda^{{\rm O}}$ in Eq.~\eqref{muEO},
we have the relation,
\begin{align}
|\mu_\lambda|^2 = \mu_\lambda\mu_\lambda^* = \left[\left(\beta^\lambda_1\right)^2 + \left(\beta^\lambda_2\right)^2\right]
\left[\left(\mu_\lambda^{{\rm E}}\right)^2 + \left(\mu_\lambda^{{\rm O}}\right)^2\right]
\approx \frac1{2k}\left[ 1 + \frac1{(k\eta)^2}\right].
\end{align}
Using this result we can obtain the power spectrum,
\begin{align}\label{PT}
{\cal P}_{{\rm T}} =\frac{64\pi G k^3}{2\pi^2}\frac{|\mu_\lambda|^2}{a^2}\approx \frac{1296 Gk^2}{\pi\alpha^6\eta^4}
\left[1+ \frac1{(k\eta)^2}\right].
\end{align}
The power spectrum of the tensor perturbation decays as the Universe
expands.\footnote{The power spectrum ${\cal P}_{{\rm T}}$
evaluated in Eq.~\eqref{PT} is valid for the whole period of the matter-dominated regime,
regardless of the background phase (contracting/expanding)
and the perturbation scale (sub-horizon/super-horizon).
[For example, when $|k\eta| \gg 2$, it reduces to that of
the initial perturbation approximated by Eq.~\eqref{hamp}.]
Therefore, apart from the bouncing point,
the amplitude of the tensor perturbation grows in time during the contracting phase,
and damps in the expanding phase.}
The power spectrum at the horizon-crossing ($|k\eta| \sim 2$) is given by
\begin{align}
{\cal P}_{{\rm T}} \approx \frac{405 Gk^6}{4\pi \alpha^6},
\end{align}
and the tensor spectral index becomes,
\begin{align}
n_{{\rm T}} \equiv \frac{d\ln {\cal P}_{{\rm T}}}{d\ln k} \approx 6.
\end{align}

\subsection{Perturbation about the Bouncing Point}
Let us discuss the tensor perturbation about the bouncing point.
The background fields, $\varphi_n(t)$, $a(t)$, and $H(t)$ had been obtained
in series expansion about $t=0$ in Ref.~\cite{Cho:2011re}.
Using the scale factor
\begin{align}
a(t) = 1 + \frac14 h_3 t^4 + \frac16 h_5 t^6 + \cdots,
\label{aaa}
\end{align}
where $h_i$'s are constants determined in Ref.~\cite{Cho:2011re},
we can transform the time coordinate $t$ to $\eta$ about the bouncing point,
\begin{align}\label{boun_time}
t &= \eta + \frac{h_3}{20} \eta^5 + \cdots.
\end{align}
Then we have the scale factor in conformal time,
\begin{align}\label{a_conf}
a = 1 + \frac14  h_3 \eta^4 + \frac16 h_5\eta^6\cdots,
\end{align}
and also have
\begin{align}\label{aa_conf}
\frac{a''}{a} = 3 h_3 \eta^2 +5 h_5 \eta^4 + \cdots.
\end{align}
With this, the field equation \eqref{tenMS} for the tensor perturbation can be solved.
Since $a''/a$ is an even function of $\eta$,
the general solution can again be expressed by a linear combination of the even and the odd mode
which are real functions,
\begin{align}\label{muEObouncing}
\mu_\lambda^{{\rm E}} &= b_0\left[1-\frac12 k^2\eta^2
+ \frac1{24}\big(k^4 +6 h_3 \big)\eta^4-\frac1{720}\big(k^6 + 42 k^2 h_3 - 120 h_5\big)\eta^6 + \cdots\right],
\nn \\
\mu_\lambda^{{\rm O}} &= b_1\left[\eta -\frac16 k^2\eta^3
+ \frac1{120}\big(k^4 + 18  h_3\big)\eta^5 - \frac1{5040}\big(k^6 + 78 k^2 h_3 - 600 h_5\big)\eta^7 + \cdots\right],
\end{align}
where $b_0$ and $b_1$ are integration constants.

In the previous subsection, we obtained the even and odd modes in the matter-dominated regime.
For a given Fourier mode, the mode coefficients are constrained by the conditions
imposed at the initial state.
Once the coefficients are fixed,
they do not change during evolution
since they are functions of the Fourier mode $k$ only.
One can see easily that the exactly same linear combination of
$\mu_\lambda^{{\rm E}}$ and $\mu_\lambda^{{\rm O}}$ in Eq.~\eqref{muEObouncing}
as in Eq.~\eqref{muTR} used for the initial perturbation,
satisfies the normalization \eqref{ten_norm}
with $b_0=b_1=\sqrt{2k}$ in the limit of $\eta\to 0$.
Therefore, one can say that
the even (odd) mode in Eq.~\eqref{muEObouncing} obtained near the bouncing point,
is continuously connected to the even (odd) mode in Eq.~\eqref{muEO} obtained
in the matter-dominated regime.
In other words, the evenness (oddness) is preserved in the whole range.
This fact will be useful when one is to solve the field equation \eqref{tenMS} numerically
from the bouncing point using Eq.~\eqref{muEObouncing}.
When one solves the field equation numerically from the bouncing point
with the even (odd) boundary conditions at $\eta=0$ using Eq.~\eqref{muEObouncing},
the numerical solution at large $|\eta|$ should reproduce the even (odd)
solution in the matter-dominated regime.
This story is valid also for the scalar perturbation in the next section.

\section{Scalar Perturbation}\label{scalar_pert}
\subsection{Initial Perturbation}
In Ref.~\cite{Cho:2011re}, we investigated the scalar perturbation.
We solved field equations numerically,
and analyzed the solutions using the even-/odd-mode technique.
In this section, we complete the analysis of the scalar perturbation
in a similar way with the tensor perturbation analyzed in the previous section.

In order to discuss the scalar perturbation,
we introduce the Sasaki-Mukhanov variable $Q_n$
which reduces to the perturbation of the scalar field
in the {\it spatially flat} gauge, $Q_n = \delta\varphi_n$.
(For details of the scalar perturbation,  see Ref.~\cite{Cho:2011re}.)
We introduce the field $v\equiv aQ_1$ for canonical quantization,
where $Q_1 = \delta\varphi_1$ corresponding to $\varphi_1$.
(For our model, the ordinary field $\varphi_1$ is dominant for the scalar perturbation,
so one may consider $v$ only.)
The action is given by
\begin{align}
S_{\rm S}= \int d\eta dx^3 \left[ \frac{1}{2} (\partial_\eta \tilde{v})^2
- \frac{1}{2} (\partial_i \tilde{v})^2 +\frac{1}{2}\frac{z''}{z}\tilde{v}^2 \right],
\end{align}
where $z=a\varphi_1'/{\cal H}$ with ${\cal H} = a'/a$.
The general solution of the equation of motion is given by
\begin{align}
\tilde{v}(\eta,\vec{x})  = \int \frac{d^3k}{(2\pi)^{3/2}}
\left[ v(\eta;k) a_{\vec{k}}  +
v^* (\eta;k) a^\dag_{-\vec{k}} \right] e^{i\vec{k}\cdot\vec{x}},
\end{align}
where $v(\eta;k)$ satisfies
\begin{align}\label{veq}
v'' + \left(k^2 - \frac{z''}{z}\right) v = 0,
\end{align}
and is normalized as
\begin{align}\label{vnorm}
v v^{*\prime} - v^* v' = i.
\end{align}

In the matter-dominated regime,
the field equation \eqref{veq} can be approximated by
\begin{align}\label{vlate}
v''(\eta) + \left[k^2 -\frac2{\eta^2} + \frac{m_1^2\alpha^6}{81} \eta^4
-\frac{4 m_1 \alpha^3}{3}\eta\,\sin \left(\frac{2m_1\alpha^3}{27} \eta^3
+ 2 \alpha_1\right)\right] v(\eta) \approx 0.
\end{align}
In the far sub-horizon limit, $|k\eta| \gg 2$,
the equation \eqref{vlate} can be further approximated by
\begin{align}
\frac{d^2v(\eta)}{d(k\eta)^2}  + \frac{m_1^2\alpha^6}{81k^6}(k\eta)^4 v(\eta) \approx 0.
\end{align}
The solution to this equation is given by
\begin{align}\label{vHankel}
v(\eta) = \sqrt{\frac{-\pi\eta}{12}}
\left[ A_1 H^{(1)}_{\frac{1}{6}} \left(-\frac{m_1\alpha^3}{27} \eta^3\right)
+ A_2 H^{(2)}_{\frac{1}{6}} \left(-\frac{m_1\alpha^3}{27} \eta^3\right) \right],
\end{align}
where $|A_2|^2 - |A_1|^2 = 1$ from normalization \eqref{vnorm}.
For $\eta \ll 0$, this solution is approximated by
\begin{align}
v(\eta) &\approx -\sqrt{\frac{9}{2m_1\alpha^3}} \;\frac{1}{\eta}\;
\left\{ A_1
\exp{\left[-i\left(\frac{m_1\alpha^3}{27}\eta^3 + \frac{\pi}{3}\right)\right]}
+ A_2
\exp{\left[i\left(\frac{m_1\alpha^3}{27}\eta^3 + \frac{\pi}{3}\right)\right]} \right\}
\\
&= -\sqrt{\frac{9}{8m_1\alpha^3}} \;\frac{1}{\eta}\; \left\{
(A_1 + A_2) \left[ \cos \left(\frac{m_1\alpha^3}{27}\eta^3 \right)
-\sqrt{3} \sin \left(\frac{m_1\alpha^3}{27}\eta^3 \right) \right]  \right.
\nn\\
& \hspace{1.2in} \left. + i  (-A_1 + A_2) \left[ \sin \left(\frac{m_1\alpha^3}{27}\eta^3 \right)
+\sqrt{3} \cos \left(\frac{m_1\alpha^3}{27}\eta^3 \right) \right]
\right\}
\\
&\equiv v^{\rm Re}(\eta) + i v^{\rm Im}(\eta).
\end{align}
If we introduce the even and the odd mode which are linearly independent real functions,
\begin{align}\label{vEO}
v^{{\rm E}} \equiv \sqrt{\frac{9}{8m_1\alpha^3}} \;\frac{1}{\eta}\; \cos\left(\frac{m_1\alpha^3}{27}\eta^3\right),
\qquad
v^{{\rm O}} \equiv -\sqrt{\frac{9}{8m_1\alpha^3}} \;\frac{1}{\eta}\; \sin\left(\frac{m_1\alpha^3}{27}\eta^3\right),
\end{align}
$v^{\rm Re}$ and $v^{\rm Im}$ can be expressed by a linear combination of
these even and odd modes.
By setting $A_1 = \beta_1 +i\beta_2$ and $A_2 = \alpha_1 +i\alpha_2$,
we obtain the following transformation,
\begin{align}
\left(
\begin{array}{c}
  v^{\rm Re} \\
 v^{\rm Im}
\end{array}
\right)
=
\left(
\begin{array}{cc}
  -\sqrt{3}\beta_1 + \beta_2 -\sqrt{3}\alpha_1-\alpha_2 & -\beta_1 -\sqrt{3}\beta_2 -\alpha_1+\sqrt{3}\alpha_2\\
  -\beta_1 -\sqrt{3}\beta_2 -\alpha_1+\sqrt{3}\alpha_2 & \sqrt{3}\beta_1 - \beta_2 -\sqrt{3}\alpha_1-\alpha_2
\end{array}
\right)
\left(
\begin{array}{c}
  v^{{\rm E}} \\
 v^{{\rm O}}
\end{array}
\right).
\end{align}
For the initial perturbation, we adopt the positive-energy mode,
$A_2=0$, i.e., $\alpha_1=\alpha_2=0$, then we have
\begin{align}\label{v_initial}
\left(
\begin{array}{c}
  v^{\rm Re}_i \\
 v^{\rm Im}_i
\end{array}
\right)
=
\left(
\begin{array}{cc}
  -\sqrt{3}\beta_1 + \beta_2 & -\beta_1 -\sqrt{3}\beta_2 \\
  -\beta_1 -\sqrt{3}\beta_2  & \sqrt{3}\beta_1 - \beta_2
\end{array}
\right)
\left(
\begin{array}{c}
  v^{{\rm E}} \\
 v^{{\rm O}}
\end{array}
\right).
\end{align}

\subsection{Super-Horizon Scale Perturbation}
In Ref.~\cite{Cho:2011re}, we observed from numerical calculations
that the scalar perturbation
gives rise to constant comoving curvature ${\cal R}$
in super-horizon scales.
In solving the field equation~\eqref{vlate} analytically,
it is not very possible to consider all terms in the equation.
For this reason, it is difficult to analyze the super-horizon scale perturbations
in terms of $v$
because they are determined by the effect of mixture of terms in Eq.~\eqref{vlate}.
Instead, it is better to analyze the super-horizon perturbations
in terms of the Sasaki-Mukhanov variable $Q_n(t)$.
The equation of motion for $Q_n$ is given by (see Ref.~\cite{Cho:2011re})
\begin{align}\label{maseq}
\ddot Q_n +3H\dot Q_n + \frac{k^2}{a^2} Q_n
+m_n^2Q_n -\frac{8\pi G}{a^3}
\sum_{l=1}^N(-1)^{l+1}\frac{d}{dt}\left(\frac{a^3}{H}
\dot\varphi_n\dot\varphi_l\right)Q_l=0.
\end{align}
When the background is relaxed to the matter-dominate universe,
we can approximate the equation for $Q_1$ which is dominant as
\begin{align} \label{lateQQ}
\ddot Q_1 + \frac{2}{t}\,\dot Q_1 +\left[\frac{k^2}{\alpha^2 t^{4/3}}
+ m_1^2-\frac{4m_1}{t}\sin (2m_1 t + 2\phi_1) \right] Q_1\approx 0.
\end{align}
The constant comoving curvature at the super-horizon scale is achieved
mainly when the $m^2$-term is dominant,
\be
\ddot Q_1 + m_1^2 Q_1 \approx 0
\quad\Rightarrow\quad
Q_1 \approx \hat A_1 e^{-im_1t} + \hat{A_2} e^{im_1t}.
\ee
If we transform this solution to the canonical field $v=\hat v (\eta)$ in conformal time,
we get
\begin{align}
\hat v(\eta) =aQ_1
&\approx  \frac{\alpha^3\eta^2}{9}
\left[ \hat A_1
\exp{\left(-i\frac{m_1\alpha^3}{27}\eta^3 \right)}
+ \hat A_2
\exp{\left(i\frac{m_1\alpha^3}{27}\eta^3 \right)} \right]
\\
&=  \frac{\alpha^3\eta^2}{9}
\left[ (\hat A_1 +\hat A_2) \cos\left(\frac{m_1\alpha^3}{27}\eta^3 \right)
+ i (-\hat A_1 +\hat A_2) \sin\left(\frac{m_1\alpha^3}{27}\eta^3 \right) \right]
\label{vtilde}
\\
&\equiv \hat v^{\rm Re}(\eta) + i \hat v^{\rm Im}(\eta).
\end{align}
If we introduce the even and the odd mode which are linearly independent,
\begin{align}
\hat v^{{\rm E}} \equiv \frac{\alpha^3\eta^2}{9} \cos\left(\frac{m_1\alpha^3}{27}\eta^3\right),
\qquad
\hat v^{{\rm O}} \equiv \frac{\alpha^3\eta^2}{9} \sin\left(\frac{m_1\alpha^3}{27}\eta^3\right),
\end{align}
$\hat v^{\rm Re}$ and $\hat v^{\rm Im}$ are expressed by a linear combination of
these even and odd modes.
By setting $\hat A_1 = \hat \beta_1 +i\hat \beta_2$ and $\hat A_2 = \hat \alpha_1 +i\hat \alpha_2$,
one obtains the following relation,
\begin{align}
\left(
\begin{array}{c}
  \hat v^{\rm Re} \\
 \hat v^{\rm Im}
\end{array}
\right)
=
\left(
\begin{array}{cc}
  \hat \beta_1 + \hat \alpha_1 & \hat \beta_2 -\hat \alpha_2\\
  \hat \beta_2 +\hat \alpha_2 & -\hat \beta_1 +\hat \alpha_1
\end{array}
\right)
\left(
\begin{array}{c}
  \hat v^{{\rm E}} \\
 \hat v^{{\rm O}}
\end{array}
\right).
\end{align}
In the previous subsection for the initial perturbation,
we adopted the positive-energy mode.
Once we pick up this mode, the positivity of the mode should remain unchanged
even in the super-horizon scale.
Therefore, we select the positive mode from the approximated solution $\hat v$
in the super-horizon scale by setting
$\hat A_2=0$ ($\hat \alpha_1=\hat \alpha_2=0$).
Then we finally get the transformation for the approximated super-horizon solutions,
\begin{align}\label{vSH}
\left(
\begin{array}{c}
  \hat v^{\rm Re} \\
 \hat v^{\rm Im}
\end{array}
\right)
=
\left(
\begin{array}{cc}
  \hat \beta_1  & \hat \beta_2 \\
  \hat \beta_2  & -\hat \beta_1
\end{array}
\right)
\left(
\begin{array}{c}
  \hat v^{{\rm E}} \\
 \hat v^{{\rm O}}
\end{array}
\right).
\end{align}

The approximated solution $\hat v(\eta)$ describes
the perturbation at the super-horizon scale for $\eta>0$
as well as for $\eta<0$.
During the contracting phase,
the initial perturbation is produced deep inside the horizon at $\eta <0$,
which is described by the solution $v(\eta)$ in Eq.~\eqref{v_initial}.
Afterwards, the perturbation crosses out the horizon and becomes the super-horizon scale.
Then, it crosses in the horizon and passes the bouncing moment.
During the expanding phase,
the perturbation crosses out the horizon soon after the bouncing moment
and becomes super-horizon scale.
Meanwhile the background universe relaxes to the matter-dominated universe,
so the super-horizon perturbation is approximated by $\hat v(\eta)$ in Eq.~\eqref{vSH}.
The power spectrum that we observe, is evaluated by this perturbation
when it crosses back into the horizon.
Afterwards, the perturbation starts to damp in principle,
and is described by $v(\eta)$ in Eq.~\eqref{v_initial} again deep inside the horizon.
However, after the perturbation crosses the horizon, it reaches the nonlinear regime
which requires modification.\footnote{The perturbation entering deep inside the
horizon at $\eta>0$ is mathematically described by the same solution $v(\eta)$
produced deep inside the horizon at $\eta<0$.
Therefore, if one solves the field equation numerically for $t>0$,
one should get the solution transiting from $\hat v$
(which provides constant comoving curvature ${\cal R}$)
to $v$ (which provides decaying ${\cal R}$).
In this sense, we misinterpreted in Ref.~\cite{Cho:2011re}
where we commented that the shape of $\hat v$
(and corresponding ${\cal R}^{\rm const}$)
maintains even after entering the horizon,
while the behavior of $v$ (and corresponding ${\cal R}^{\rm decaying} \propto t^{-1}$)
is sub-dominantly implied in $\hat v$ in the sub-horizon scales.}

The approximated super-horizon solution $\hat v(\eta)$ in Eq.~\eqref{vtilde} is not normalizable,
i.e., it does not satisfy the normalization condition \eqref{vnorm}.
The correct normalizable super-horizon solution $\check v(\eta)$
must require some corrections to the approximated solution.
However, it is not very possible to obtain $\check v(\eta)$ in our situation.
Therefore, let us evaluate the power spectrum in approximation using $\hat v$,
which will imply underlying physics to some extent.

The power spectrum for the scalar perturbation is given by
\begin{align}
{\cal P}_{\cal R} = \frac{k^3}{2\pi^2}|{\cal R}|^2,
\label{PS}
\end{align}
where the {\it comoving curvature} is defined by
\begin{align}\label{RR}
{\cal R} = \frac{H}{\dot\varphi_1^2-\dot\varphi_2^2}
\left(\dot\varphi_1 Q_1 - \dot\varphi_2 Q_2 \right)
\equiv f_1 Q_1 - f_2 Q_2 \approx f_1Q_1.
\end{align}
In the last step, we used the fact that $Q_1$ is dominant.
When the background is relaxed to the matter-dominated universe,
using Eq.~\eqref{BGfields} we have
\begin{align}\label{f1}
f_1 = \frac{H}{\dot\varphi_1^2-\dot\varphi_2^2}\dot\varphi_1
= -4\pi G \frac{H}{\dot H}\dot\varphi_1
\approx 4\pi G t \dot\varphi_1
= \sqrt{\frac{16\pi G}{3}} \cos\left(\frac{m_1\alpha^3}{27}\eta^3\right).
\end{align}
The approximated comoving curvature is the given by
\begin{align}
\hat {\cal R}  = \frac{f_a}{a} \hat v,
\end{align}
then the approximated power spectrum which is averaged over time, becomes
\begin{align}
\left\langle {\cal P}_{\hat {\cal R}} \right\rangle
= \left\langle \frac{k^3}{2\pi^2} |\hat {\cal R}|^2 \right\rangle
= \left\langle \frac{k^3}{2\pi^2} \frac{f_1^2}{a^2} \left[ \left(\hat v^{\rm Re} \right)^2
+ \left(\hat v^{\rm Im} \right)^2 \right] \right\rangle
\approx \frac{16 G}{3\pi} k^3 {\cal B}^2(k) ,
\end{align}
where ${\cal B}^2(k) = \hat\beta_1^2 + \hat\beta_2^2$
and the cosine factor has been averaged out.
Unfortunately since $\hat v$ does not satisfy the normalization condition \eqref{vnorm},
we do not know the value of $\hat\beta_i$.
In order to provide the scale-invariant power spectrum,
${\cal B}^2(k) \propto k^{-3}$ is required.\footnote{If we consider
$|{\cal R}|^2$ for far sub-horizon scales,
$f_1$ is the same with Eq.~\eqref{f1},
and we can use the solutions $v_i^{\rm Re}$ and $v_i^{\rm Im}$ in Eq.~\eqref{v_initial}
evaluated by $v^{\rm E}$ and $v^{\rm O}$ in Eq.~\eqref{vEO}.
Then we have
\begin{align}
\left\langle |{\cal R}|^2 \right\rangle \propto
\left\langle \frac{f_1^2}{a^2} \left[ \left( v^{\rm Re} \right)^2
+ \left( v^{\rm Im} \right)^2 \right] \right\rangle
\propto \frac{1}{\eta^6} .
\end{align}
Therefore, $\left\langle |{\cal R}|^2 \right\rangle$ grows in the contracting phase,
and decays in the expanding phase.}

\section{Conclusions}
In this paper, we studied the tensor and scalar perturbations of the symmetric Lee-Wick bouncing universe.
We adopted the even-/odd-mode analysis instead of decaying-/constant-mode analysis
in the literature~\cite{Allen:2004vz,Kim:2006ju,Cai:2008qw}.
The even and the odd mode are linearly independent modes of the perturbation.
The analytic solutions for these modes are obtained both in the bouncing point
and the matter-dominated regime.
The evenness and the oddness are preserved in these two regimes.

For the tensor perturbation, the mode functions are analytically solved
when the background is in the matter-dominated universe.
These mode functions are valid both for the sub-horizon and the super-horizon scales.
The initial tensor perturbation produced deep inside the horizon
during the contracting phase at $t<0$, grows in time.
After the tensor mode crosses out the horizon during the contracting phase,
it grows more rapidly.
After the bouncing, the perturbation evolves in a symmetric way;
it decays as the Universe expands.
In the super-horizon scales during the expanding phase,
the power spectrum ${\cal P}_{\rm T}$ decays in time.
Finally when it crosses the horizon,
the spectral index is evaluated as $n_{\rm T} =6$.

For the scalar perturbation,
the initial perturbation obtained in Ref.~\cite{Cho:2011re}
provides growing comoving curvature ${\cal R}$.
In the super-horizon scales, we obtained the approximated solutions
for the even and odd modes.
These approximated solutions provide constant comoving curvature.
However, they are not complete enough to evaluate the $k$-dependence of the power spectrum.
(The approximated mode solution is not normalizable,
so the mode coefficients which depend on $k$ are not determined.
The correct mode solution would require some corrections to this approximate solutions,
but is difficult to obtain for our model.)
Therefore, the scale invariance could not be examined.

When the scalar perturbation finally crosses in the horizon during the expanding phase,
the modes are of the same functional form with the initial perturbation due to symmetry.
The corresponding comoving curvature will decay, and
the perturbation evolves finally into the nonlinear regime which is beyond our scope.
As a whole, the comoving curvature by the scalar perturbation
produced initially in the sub-horizon at $t<0$,
grows to become constant in the super-horizon.
During the expanding phase at $t>0$,
the curvature remains constant in the super-horizon, and starts to decay after
the perturbation crosses into the horizon.

\subsection*{Acknowledgements}
We are grateful to Jinn-Ouk Gong, Seoktae Koh,
and Jai-chan Hwang for helpful discussions.
This work was supported by the Korea Research Foundation (KRF) grants
funded by the Korea government (MEST) No. 2009-0070303 and No. 2012-006136 (I.C.),
and No. 2011-0009972 (O.K.).

\clearpage
\begin{figure}
\centerline{\epsfig{figure=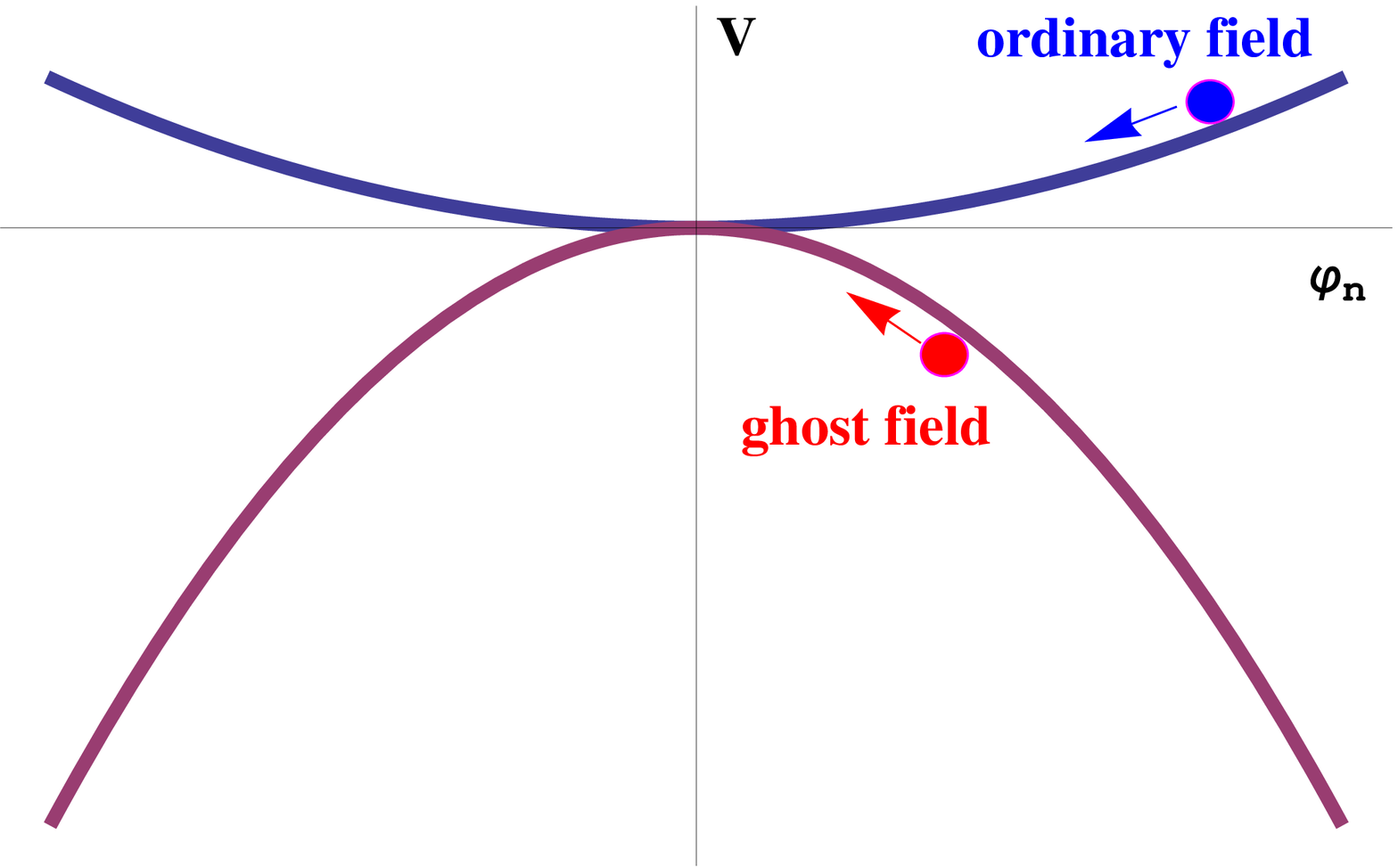,height=70mm}}
\caption{
At the bouncing point ($t=0$),
the ordinary field $\varphi_1$ and the ghost field $\varphi_2$
are at rest, $\dot\varphi_n=0$ ,
with the values satisfying $m_1^2\varphi_1^2 = m_2^2\varphi_2^2$.
Afterwards at $t>0$, two fields approach the center of the potential,
and start to oscillate about $\varphi_n=0$.
For $t<0$, the behavior of $\varphi_n$ is evenly symmetric.
For $t>0$, the background universe expands due to two scalar fields,
approaching the matter-dominated universe.
For $t<0$, the universe contracts in a symmetrical way.
}
\label{FIG1}
\end{figure}
\begin{figure}
\centerline{\epsfig{figure=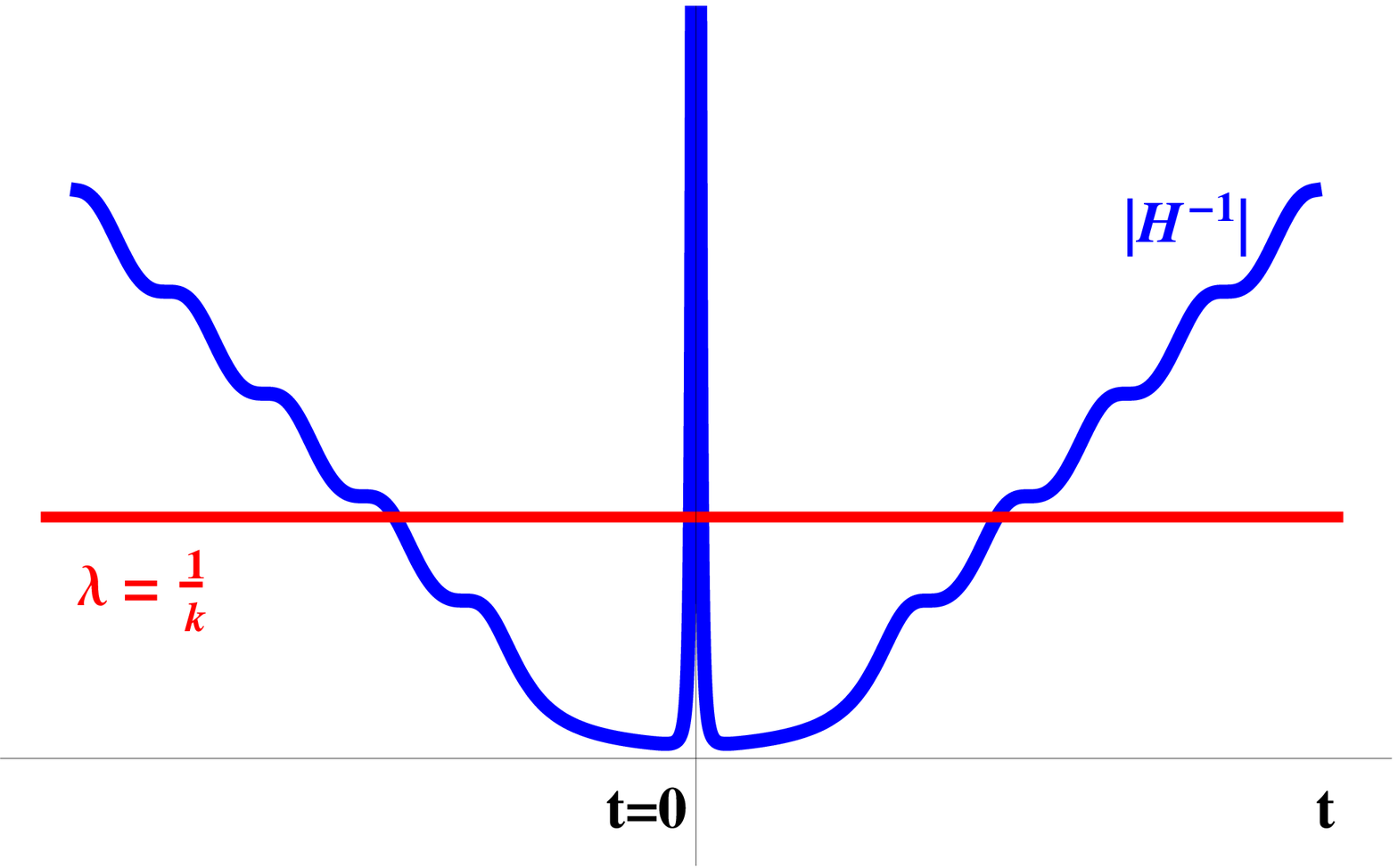,height=70mm}}
\caption{
Plot of the horizon scale $|H^{-1}|$.
The perturbation for a given wave length is initially produced
inside the horizon during the contracting phase at $t<0$,
and crosses the horizon four times.
}
\label{FIG2}
\end{figure}


\begin{thebibliography}{99}

\bibitem{Lee:1969fy}
  T.~D.~Lee and G.~C.~Wick,
  ``Negative Metric and the Unitarity of the S Matrix,''
  Nucl.\ Phys.\  B {\bf 9}, 209 (1969);
  ``Finite Theory of Quantum Electrodynamics,''
  Phys.\ Rev.\  D {\bf 2}, 1033 (1970).


\bibitem{Grinstein:2007mp}
  B.~Grinstein, D.~O'Connell and M.~B.~Wise,
  ``The Lee-Wick standard model,''
  Phys.\ Rev.\  D {\bf 77}, 025012 (2008)
  [arXiv:0704.1845 [hep-ph]].


\bibitem{Carone:2008iw}
  C.~D.~Carone and R.~F.~Lebed,
  ``A Higher-Derivative Lee-Wick Standard Model,''
  JHEP {\bf 0901}, 043 (2009)
  [arXiv:0811.4150 [hep-ph]];

  R.~F.~Lebed and R.~H.~TerBeek,
  ``Collider Signatures of the N=3 Lee-Wick Standard Model,''
  JHEP {\bf 1209}, 099 (2012)
  [arXiv:1205.3213 [hep-ph]].

\bibitem{Cho:2010hj}
  I.~Cho and O.~K.~Kwon,
  ``Generalized Lee-Wick Formulation from Higher Derivative Field Theories,''
  Phys.\ Rev.\  D {\bf 82}, 025013 (2010)
  [arXiv:1003.2716 [hep-th]].

\bibitem{Grinstein:2007iz}
  B.~Grinstein, D.~O'Connell and M.~B.~Wise,
  ``Massive Vector Scattering in Lee-Wick Gauge Theory,''
  Phys.\ Rev.\  D {\bf 77}, 065010 (2008)
  [arXiv:0710.5528 [hep-ph]];
  ``Causality as an emergent macroscopic phenomenon: The Lee-Wick O(N) model,''
  Phys.\ Rev.\  D {\bf 79}, 105019 (2009)
  [arXiv:0805.2156 [hep-th]].



\bibitem{Fornal:2009xc}
  B.~Fornal, B.~Grinstein and M.~B.~Wise,
  ``Lee-Wick Theories at High Temperature,''
  Phys.\ Lett.\ B {\bf 674}, 330 (2009)
  [arXiv:0902.1585 [hep-th]].

\bibitem{Accioly:2011zz}
  A.~Accioly, P.~Gaete, J.~Helayel-Neto, E.~Scatena and R.~Turcati,
  ``Investigations in the Lee-Wick electrodynamics,''
  Mod.\ Phys.\ Lett.\ A {\bf 26}, 1985 (2011).



\bibitem{Khoury:2001wf}
  J.~Khoury, B.~A.~Ovrut, P.~J.~Steinhardt and N.~Turok,
  ``The Ekpyrotic universe: Colliding branes and the origin of the hot big bang,''
  Phys.\ Rev.\ D {\bf 64}, 123522 (2001)
  [hep-th/0103239].





\bibitem{Cai:2008qw}
  Y.~F.~Cai, T.~t.~Qiu, R.~Brandenberger and X.~m.~Zhang,
  ``A Nonsingular Cosmology with a Scale-Invariant Spectrum of Cosmological
  Perturbations from Lee-Wick Theory,''
  Phys.\ Rev.\  D {\bf 80}, 023511 (2009)
  [arXiv:0810.4677 [hep-th]].

\bibitem{Cai:2009fn}
  Y.~F.~Cai, W.~Xue, R.~Brandenberger and X.~Zhang,
  ``Non-Gaussianity in a Matter Bounce,''
  JCAP {\bf 0905}, 011 (2009)
  [arXiv:0903.0631 [astro-ph.CO]];

  J.~Karouby and R.~Brandenberger,
  ``A Radiation Bounce from the Lee-Wick Construction?,''
  Phys.\ Rev.\  D {\bf 82}, 063532 (2010)
  [arXiv:1004.4947 [hep-th]];

  J.~Karouby, T.~Qiu and R.~Brandenberger,
  ``On the Instability of the Lee-Wick Bounce,''
  Phys.\ Rev.\  D {\bf 84}, 043505 (2011)
  [arXiv:1104.3193 [hep-th]];

  Y.~F.~Cai, R.~Brandenberger and X.~Zhang,
  ``Preheating a bouncing universe,''
  Phys.\ Lett.\  B {\bf 703}, 25 (2011)
  [arXiv:1105.4286 [hep-th]].

 \bibitem{Zhang:2010bb}
  J.~Zhang, Z.~G.~Liu and Y.~S.~Piao,
  ``Amplification of curvature perturbations in cyclic cosmology,''
  Phys.\ Rev.\  D {\bf 82}, 123505 (2010)
  [arXiv:1007.2498 [hep-th]];


  B.~Xue and P.~J.~Steinhardt,
  ``Evolution of curvature and anisotropy near a nonsingular bounce,''
  Phys.\ Rev.\ D {\bf 84}, 083520 (2011)
  [arXiv:1106.1416 [hep-th]];

  D.~A.~Easson, I.~Sawicki and A.~Vikman,
  ``G-Bounce,''
  JCAP {\bf 1111}, 021 (2011)
  [arXiv:1109.1047 [hep-th]].


\bibitem{Cho:2011re}
  I.~Cho and O-K.~Kwon,
  ``Scalar Perturbation in Symmetric Lee-Wick Bouncing Universe,''
  JCAP {\bf 1111}, 043 (2011)
  [arXiv:1109.5753 [gr-qc]].

\bibitem{Allen:2004vz}
  L.~E.~Allen and D.~Wands,
  ``Cosmological perturbations through a simple bounce,''
  Phys.\ Rev.\  D {\bf 70}, 063515 (2004)
  [arXiv:astro-ph/0404441].

\bibitem{Kim:2006ju}
  H.~S.~Kim and J.~c.~Hwang,
  ``Evolution of linear perturbations through a bouncing world model: Is the
  Harrison-Zel'dovich spectrum possible via bounce?,''
  Phys.\ Rev.\  D {\bf 75}, 043501 (2007)
  [arXiv:astro-ph/0607464].

\end{thebibliography}
\end{document}